\documentclass[prd,preprint,amsmath,nofootinbib,superscriptaddress]{revtex4}
\usepackage{graphicx}
\usepackage{bm}
\usepackage{epsfig}

\newcommand{\beq}{\begin{equation}}
\newcommand{\eeq}{\end{equation}}
\newcommand{\bea}{\begin{eqnarray}}
\newcommand{\eea}{\end{eqnarray}}

\begin{document}
\title{Towers of Gravitational Theories} 
\author{Walter D. Goldberger\footnote{\uppercase{W}ork supported by \uppercase{DOE}
contract \uppercase{DE-FG}02-92ER40704.
\uppercase{DEAC02-6CH03000}. Email address walter.goldberger@yale.edu}}
\address{Dept. of Physics, Yale University, New Haven CT 06520, USA}

\author{Ira Z. Rothstein\footnote{\uppercase{W}ork supported by \uppercase{DOE}
contracts \uppercase{DOE-ER}-40682-143 and
\uppercase{DEAC02-6CH03000}. Email address izr@andrew.cmu.edu}}
\address{Dept. of Physics, Carnegie Mellon University, 
Pittsburgh PA  15213, USA}
\begin{abstract}{In  this essay we  introduce a theoretical framework designed to 
describe  black hole dynamics.  The difficulties in understanding such dynamics stems from the proliferation of scales involved when one attempts to simultaneously describe all of the relevant  dynamical degrees of freedom.  These range  from the modes that describe the black hole horizon, which are responsible for dissipative effects, to the long wavelength gravitational radiation that drains mechanical energy from macroscopic black hole bound states.  We approach the problem from a Wilsonian point of view, by building a tower of theories of gravity each of which is valid at different scales. The methodology leads to multiple new results in diverse topics including phase transitions of Kaluza-Klein black holes and the interactions of spinning black hole in non-relativistic orbits.  Moreover, our methods tie together speculative ideas regarding dualities for black hole horizons to real physical measurements in gravitational wave detectors.
}
\end{abstract}
\maketitle
\newpage

The current experimental efforts in gravitational wave detection have brought renewed impetus to the problem of predicting the motion of extended gravitationally bound objects within the context of general relativity.  This ``problem of motion'' is one of the most fundamental and vexing questions in physics, whose complexity owes itself in part to the presence of several dynamical length scales that arise in any attempt at a solution.  The challenges involved may be brought to light by breaking the problem down into two sub-problems, namely the motion of the internal degrees of freedom of each body, and the motion of each body's center of mass.  Of course, these two problems are coupled since, for instance, the gravitational field of one body will exert tidal forces on the other object which in turn will deform its shape, cause it to spin, and accelerate its center of mass.  In addition, one must account for the fact that as each object accelerates, it emits and absorbs gravitational radiation which itself backreacts on its 
motion.  Understanding the evolution of this coupled systems in an analytically controlled manner is quickly seen to be an intractable problem, except perhaps in specialized situations that have either a high degree of symmetry or a suitable expansion parameter (for instance the relative velocity, in the case of nearly Newtonian orbits).  

The simplest version of the two-body problem occurs when the gravitationally bound system consists of a pair of black holes.  In this situation the internal as well as the external forces are strictly gravitational in origin, and the physics is well described by pure general relativity.  The ``internal problem" is then the problem of understanding the dynamics of the black holes themselves.   At a fundamental level, the internal dynamics is characterized by a spectrum of resonant states, or quasinormal modes (QNM's).   These modes, which can be pictured as the ``ringing" of a black hole spacetime in response to external perturbations, contain all information about classical processes involving isolated black holes, including for instance scattering and absorption cross sections of low energy gravitons.  

Since the same degrees of freedom that describe the interactions of black holes with gravitons also govern their interactions with other black holes, the pattern of gravitational radiation emitted by a binary black hole system contains in it the imprints of the structure of the individual black holes. Thus the data obtained by detectors such as LIGO or LISA could in principle be used to answer questions regarding some of the simplest yet most mysterious objects in nature, and to test our understanding of ideas in general relativity that have up to now only been realized in the realm of gedanken experiments.  However, in order to extract this physics, one needs to have a method for disentangling from the data the short distance signatures of black holes from the well-understood long wavelength physics.

Our approach to the problem is Wilsonian in nature. That is, we develop a tower of effective field theories
in such a way that we may cleanly separate out the physics at disparate scales~\cite{GnR1}.  The shortest scales in the problem are those which dictate the  ``internal dynamics of the body''. In the case of  black hole these would be the QNM (we will be referring to black holes from now on unless otherwise stated).  Upon integrating out internal modes, whose wavelength is of order the Schwarzschild radius,  one is left with a theory  which we call the point particle effective theory (PPEFT),  described by an  action of the form
\begin{equation}
\label{LPPEFT}
S_{eff}=-m \int d\tau+c_E \int d\tau E_{\mu \nu}E^{\mu \nu}+c_B \int d\tau B_{\mu \nu} B^{\mu \nu}  +\cdots
\end{equation}
Where $E_{\mu\nu}$ and $B_{\mu \nu}$ denote the decomposition of the Weyl tensor in terms of electric and magnetic type parity respectively.  In addition to the operators, shown here, there are in general an infinite set of operators with more derivatives acting on the gravitational field.   Operators constructed from the Ricci tensor can be removed by field redefinitions, so the infinite tower of operators simply involves powers of $E_{\mu\nu}$ and $B_{\mu\nu}$.

For this theory to have predictive power one must have a small parameter that allows the truncation of the expansion in Eq.~({\ref{LPPEFT}}).   The necessary expansion parameter is given by  $r_s/{\cal R} \ll 1$ where $r_s$ is the black hole radius and ${\cal R}$ is a scale that characterizes the gradients of the gravitational field.  In the examples we will be considering ${\cal R}$ will correspond to either the compactifaction scale of an internal manifold, or to the orbital radius in a black hole binary.  Note that the coefficients of operators in $S_{eff}$ can be fixed by a matching calculation of the same sort employed in any effective quantum field theory.   One simply adjusts the coefficients $c_{E,B},\cdots$ in such a way that Eq.~(\ref{LPPEFT}) reproduces observables in the ``full theory" consisting of an isolated black hole.   The operatiors in $S_{eff}$ beyond the usual kinetic term account for finite size effects.   By including all such terms at a given order in $r_s/{\cal R}$, one can systematically account for the internal structure of the black hole.

It is well known that calculations in general relativity coupled to point sources exhibit short distance singularities.  In the past this has lead to conceptual stumbling blocks.  But in the PPEFT, these divergence are handled quite naturally.  All divergences  can be absorbed into counterterms for the
higher dimensional operators.  Moreover, these divergences induce  non-trivial renormalization
group flows, and all of the standard quantum field theoretic tools can be applied.  

Before integrating discussing the appropriate EFT in the multi-black hole sector, which can be obtained by integrating out modes in PPEFT,  we give here a simple application of $S_{eff}$ in Eq.~(\ref{LPPEFT}).   Consider a Kaluza-Klein black hole, that is, a black hole embedded in a space of the form ${\bf R}^{d-1}\times S^1$.   For different regimes of the parameter $\lambda\equiv r_s/L$, with $L$ the asymptotic radius of the $S^1$ direction, this system is believed to undergo phase transitions connecting black objects of different horizon topology.   This setup is of interest because it serves as a toy model of topology change in higher dimensional gravity, for its possible connections, through gauge/gravity duality,  to phase transitions finite temperature Yang-Mills theory in the large $N$ limit~\cite{Harmark}, and for their possible role in high energy collisions in models of low scale gravity.

\begin{figure*}[t!]
\def\size{5cm}
\hbox{\vbox{\hbox to \size {\hfil \includegraphics[width=4cm]{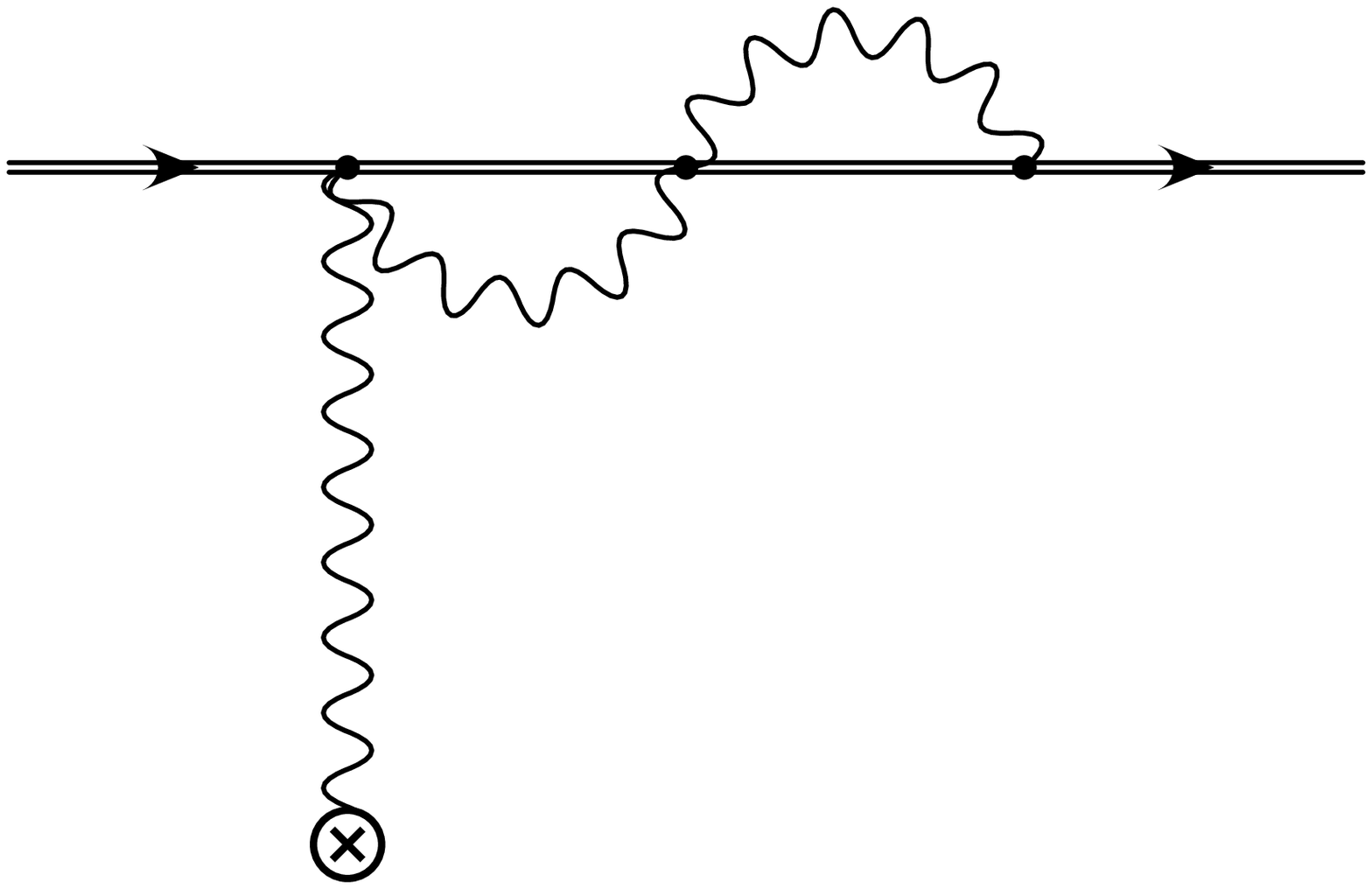} \hfil }\hbox to \size {\hfil(a)\hfil}}
\vbox{\hbox to \size {\hfil \includegraphics[width=4cm]{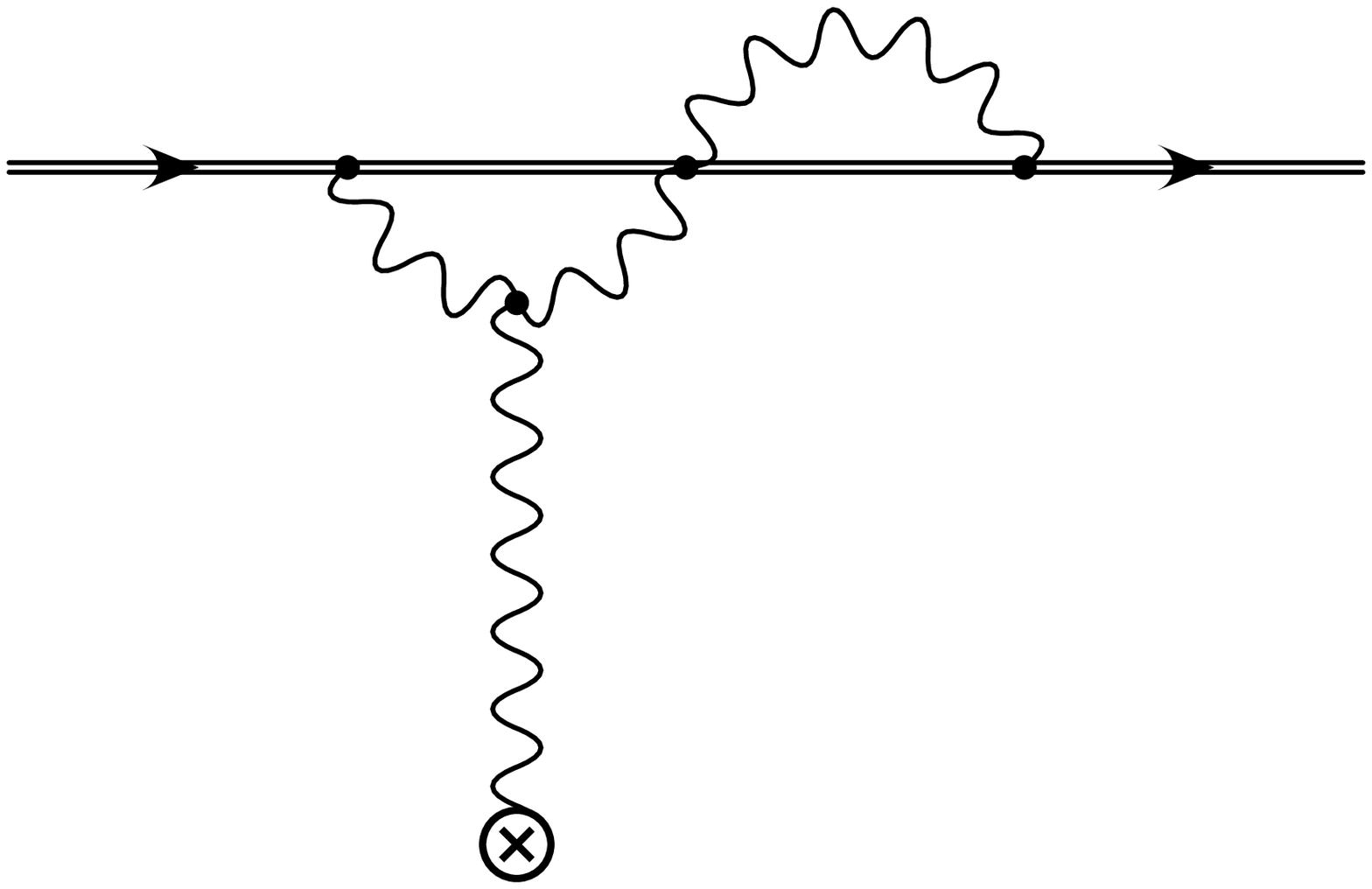} \hfil}\hbox to \size {\hfil(b)\hfil}}
\vbox{\hbox to \size {\hfil \includegraphics[width=4cm]{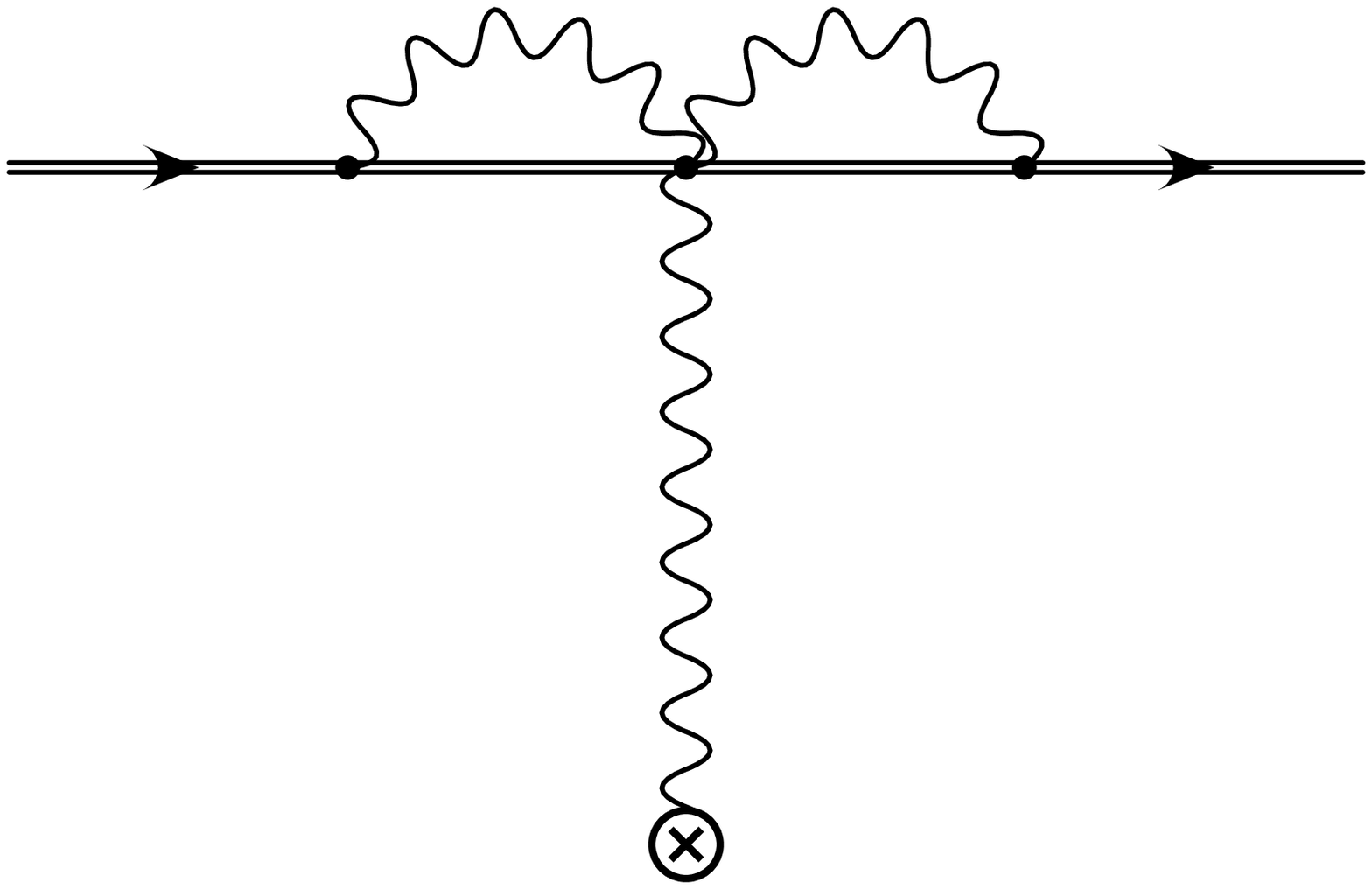} \hfil}\hbox to \size {\hfil(c)\hfil}}}
\hbox{\vbox{\hbox to \size {\hfil \includegraphics[width=4cm]{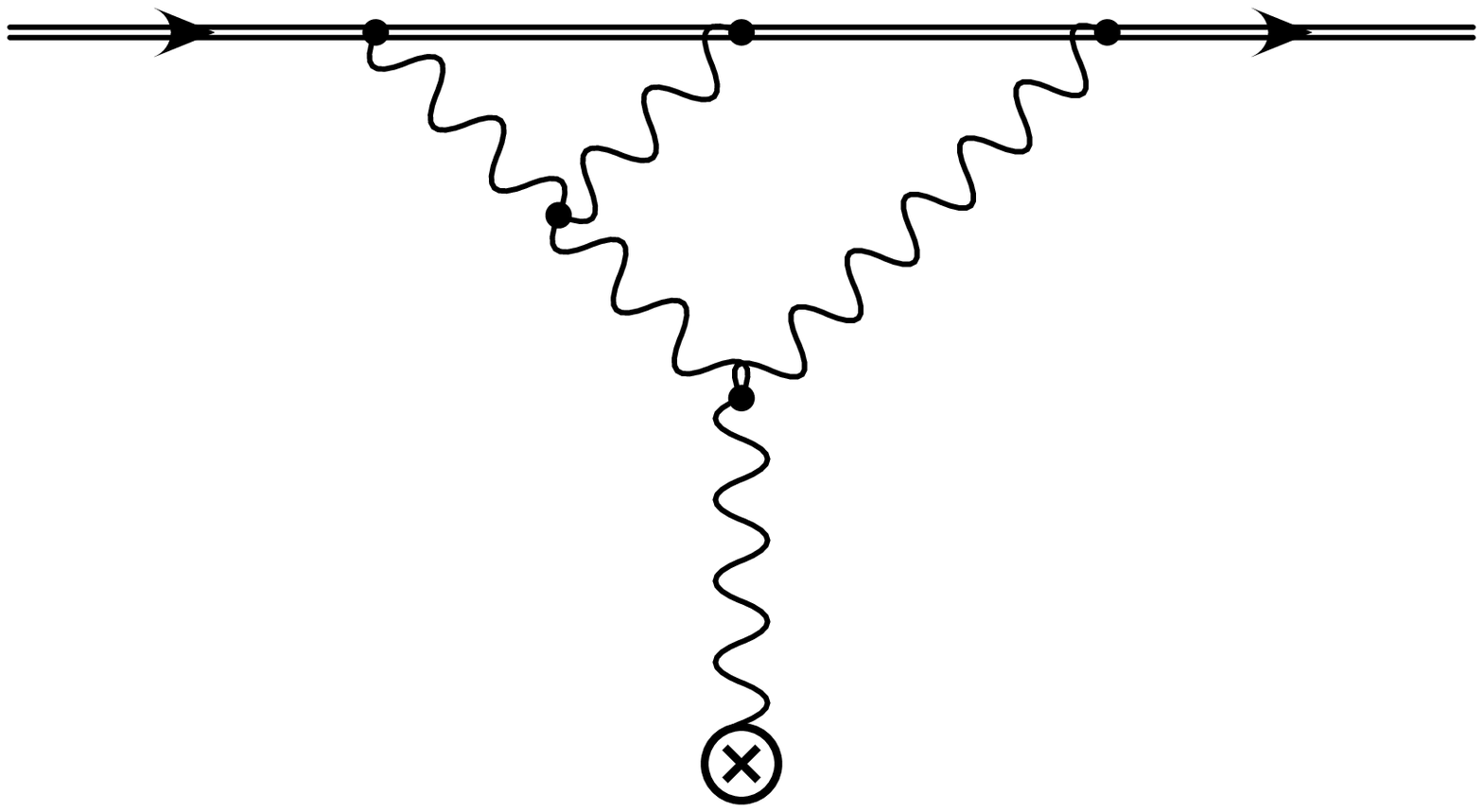} \hfil }\hbox to \size {\hfil(d)\hfil}}
\vbox{\hbox to \size {\hfil \includegraphics[width=4cm]{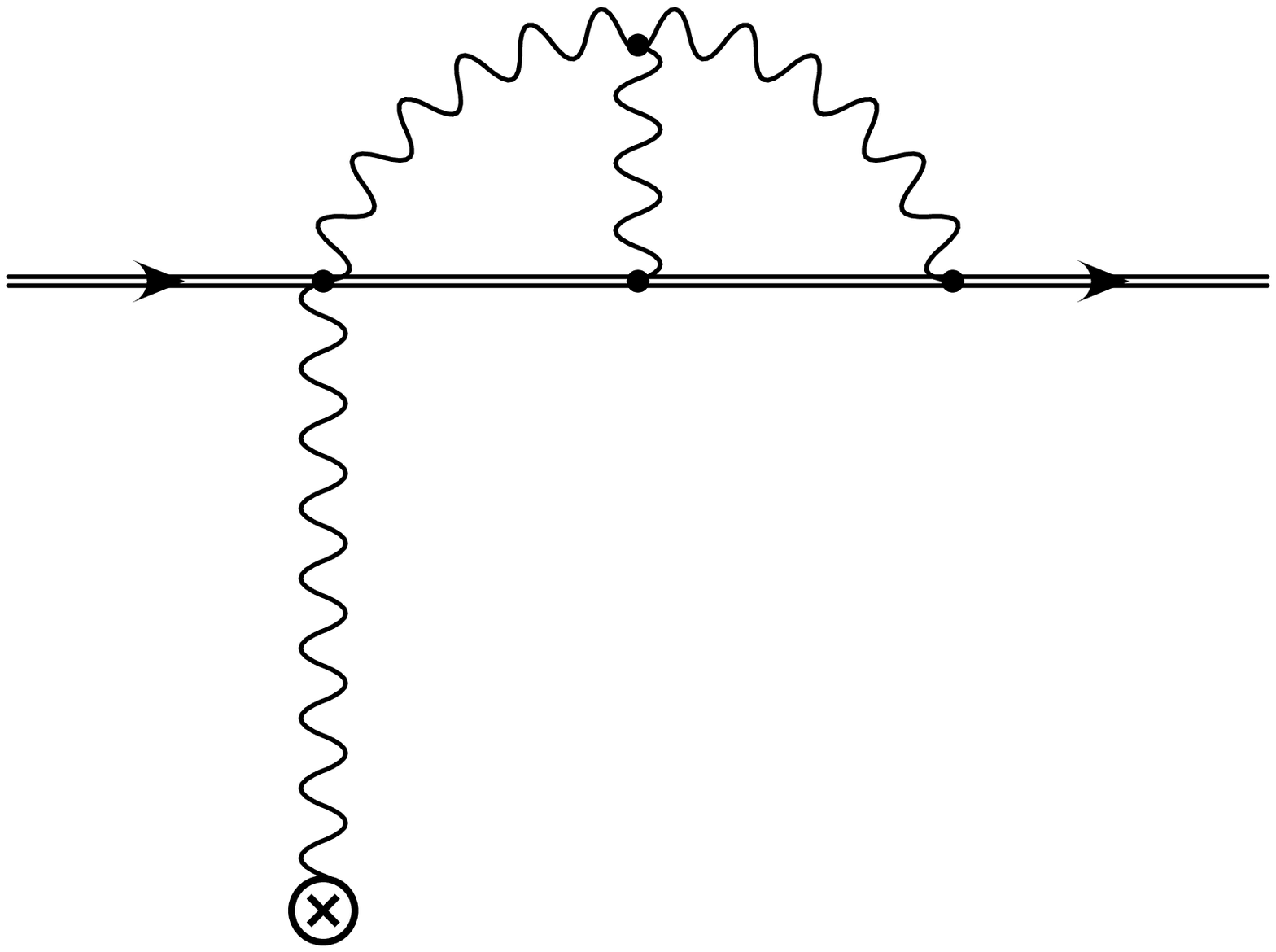} \hfil}\hbox to \size {\hfil(e)\hfil}}
\vbox{\hbox to \size {\hfil \includegraphics[width=4cm]{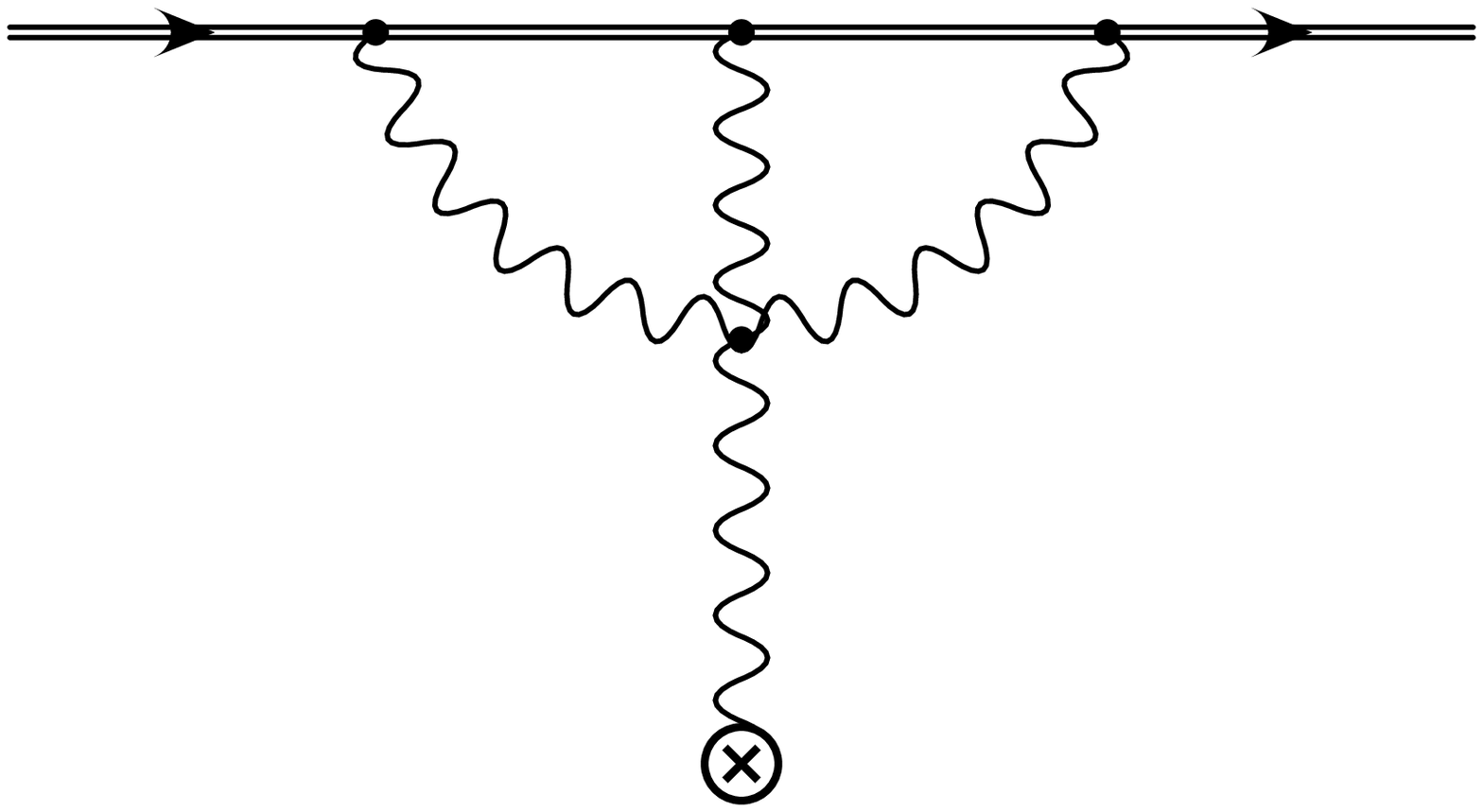} \hfil}\hbox to \size {\hfil(f)\hfil}}}
\caption{Diagrams contributing to the mass and tension at order $\lambda^2$.}
\label{m3}
\end{figure*}

If we imagine starting with a black string whose horizon wraps the $S^1$ direction,  then as $\lambda$ approaches one, the string can decay due to the Gregory-Laflamme instability~\cite{GL}.  The phase diagram for the thermodynamics of this transition has been studied numerically for the full range of $\lambda$, and analytically to leading order in $\lambda$ in the limit where $\lambda\ll 1$.  PPEFT allows one to calculate systemically by starting in the point particle approximation and then adding finite size effects via the contribution of the higher dimensional operators in Eq.~(\ref{LPPEFT}).   The calculation involves the computation of the asymptotic graviton tadpole $\langle h_{\mu\nu}\rangle$ from which the black hole mass $m$ and tension $\tau$ can be read off.   From these two quantitites all the black hole thermodynamics can be reconstructed.    For $\lambda\ll 1$, the tadpole $\langle h_{\mu\nu}\rangle$ has a natural expansion in terms of standard Feynman diagrams.   These diagrams follow from the Feynman rules implied by Eq.~(\ref{LPPEFT}).   Any divergences that arise due to the point particle limit can be easily handled by calculating the diagrams in dimensional regularization.   Formulating the problem in this way, we have obtained~\cite{Chu} the thermodynamic quantities to order $\lambda^2$, extending the known results in $d=5$ to arbitray dimension.   The diagrams contributing to the order $\lambda^2$ thermodynamics are shown in Fig.~\ref{m3}.

Given the mass $m$ and the tension $\tau$ obtained from the calculation of the one-point function, all other thermodynamic quantites can be easily obtained.    One simply uses the Smarr relation in $d$ dimensions.
\begin{equation}
TS =\frac{d-3}{d-2}m-\frac{1}{d-2} \tau L.
\end{equation}
The results are
\begin{equation}
{S\over S(L\rightarrow\infty)} = 1 +{1\over 2} \left({d-2\over d-3}\right) \zeta(d-3) {2 G_N m\over L^{d-3}} + {1\over 8} {d-2\over (d-3)^2} \zeta^2(d-3) \left({2 G_N m\over L^{d-3}}\right)^2 +\cdots,
\end{equation}
and 
\begin{equation}
{T\over T(L\rightarrow\infty)} = 1 - {2d-5\over 2(d-3)}\zeta(d-3) {2 G_N m\over L^{d-3}} +\left[8 d^2 - 43 d + 58\over 8(d-3)^2\right]\zeta^2(d-3) \left({2 G_N m\over L^{d-3}}\right)^2+\cdots,
\end{equation}
where $S(L\rightarrow\infty)$ and $T(L\rightarrow\infty)$ are the entropy and temperature of an uncompactified black hole.   These result were compared in~\cite{Chu} to numerical calculations~\cite{Kudoh} of the phase diagram over the full range of $\lambda$ for $d=5,6$.  At values of $\lambda$ for which the analytic calculations can be trusted there seems to be a relevant discrepancy between our results and the numerical data.  However, no conclusion can be reached until an estimate of errors in the numerical data is given~\footnote{Note that our results, agree with known order $\lambda^2$ results for $d=5$~\cite{Karasik} done by black hole perturbation theory methods, providing a check of our more general calculation.}.   The main conclusion to be drawn from our analysis is that, according to the power counting PPEFT, finite size effects characteristic of a black hole do not arise until order $\lambda^{2(d-1)/(d-3)}$.  Therefore despite of using the machinery of black hole perturbation theory in intermediate steps,  none of the analytic results achieved to date are sensitive to the details of the black hole structure.

A more physically relevant example of the uses of Wilsonian methods in gravitational physics involves the dynamics of black hole binary bound states.   These systems are important for the experimental program in gravitational wave detection in LIGO/VIRGO and LISA, as they constitute the binary inspiral signal in the non-relativistic regime.  The binary system involves multiple scales beyond those in the previous example:  besides the typical black hole radius $r_s$, there are new scales $r$, the orbital length, and $r/v$, the wavelength of gravitons emitted from the binary.   Note that by the virial theorem, these scales are all correlated, for instance $r\sim r_s/v^2$ (here $v\ll 1$ is the typical orbital velocity, which serves as the small expansion parameter\footnote{Another possible expansion parameter, relevant for LISA physics, arises if the binary constituents have mass ratios that differ significantly from unity.   The limit $m/M\ll 1$ can also be handled systematically within PPEFT.} in the multi-black hole EFT).

To treat the binary black hole problem, one must go beyond the PPEFT.  To do se one simply integrates out all modes of the graviton with wavelengths between the scales $r_s$ and $r$.   The result of this procedure is an effective Lagrangian describing the interactions of a composite particle (roughly the center of mass coordinate of the binary) with time dependent moments that interact with the long wavelength modes of the gravitational field.    Explicitly, this is done by decomposing 
the metric  in terms of a short wavelength ``potential graviton" $H_{\mu\nu}$ and a long wavelength background field  ${\bar g}_{\mu\nu}$ which will ultimately reproduce the effects of radiation,
\begin{equation}
g_{\mu \nu}={\bar g}_{\mu \nu}+H_{\mu \nu}.
\end{equation}
Working in the background field gauge in order to preserve gauge invariance under transformations of the ${\bar g}_{\mu\nu}$, the effective action is formally given by
\begin{equation}
\exp[i\Gamma(x_i,{\bar g})]=\int d H_{\mu\nu}(x) \exp\left(i S_{eff}[x_i,{\bar g}+H] +iS_{EH}[\bar{g}+H]\right),
\end{equation}
where $x_i$ denotes the worldline of the $i$-th black hole and $S_{EH}=-2 m^2_{Pl}\int d^4 x \sqrt{g} R$ is the Einstein-Hilbert term.   The functional integration is performed in practice by calculating all Feynman diagrams which are one-particle irreducible and involve only the radiation graviton ${\bar h}_{\mu\nu}={\bar g}_{\mu\nu}-\eta_{\mu\nu}$ in external states.    The result of doing this includes all post-Newtonian corrections to the two particle Lagrangian in the zero radiation graviton sector, for example at  order $v^2$ beyond the Newtonian potential, one finds in background field gauge,
\begin{eqnarray}
\label{EIH}
\nonumber
L_{EIH} &=& {1\over 8}\sum_a m_a {\bf v}^4_a + {G_N m_1 m_2\over 2 |{\bf x}_1 -{\bf x}_2|}\left[3 ({\bf v}^2_1+ {\bf v}^2_2) - 7({\bf v}_1\cdot {\bf v}_2)  - {({\bf v}_1\cdot  {\bf x}_{12} )({\bf v}_2\cdot {\bf x}_{12})\over |{\bf x}_1-{\bf x}_2|^2} \right]\\
& & {} - {G^2_N m_1 m_2 (m_1+m_2)\over 2  |{\bf x}_1 -{\bf x}_2|^2},
\end{eqnarray}
first derived by Eintein, Infeld and Hoffman.     In the one-graviton sector, the result of carrying out the functional integration gives rise to 
\begin{eqnarray}
\label{eq:NRGRv2}
\Gamma_{v^{5/2}} &=& -{{ h}_{00} \over 2 m_{Pl}}  \left[{1\over 2}\sum_a m_a {\bf v}^2_a -{G_N m_1 m_2\over |{\bf x}_1 -{\bf x}_2|}\right]\\
\nonumber
& &  -{1\over 2 m_{Pl}}\epsilon_{ijk} {\bf L}_k \partial_j { h}_{i0} + {1\over 2 m_{Pl}}\sum_a m_a {{\bf x}_a}_i {{\bf x}_a}_j R_{0i0j}.
\end{eqnarray}
Integrating out the multipole expanded \cite{GR} radiation modes gives a non-local effective Lagrangian for the center of mass coordinate of the binary star.   This effective Lagrangian has an imaginary part, which at leading order is due to a Feynman a diagram with two insertions of $\Gamma_{v^{5/2}}$.    This imaginary part signifies an instability, which is none other than the power lost by the binary system due to the emission of gravitational radiation.   Effects such as radiation reaction are also naturally incorporated in this effectve theory.    For instance radiation reaction corresponds to a non-local term in the effective Lagrangian.   To leading order this is simply the gravitational analog of the Abraham-Lorentz-Dirac equation.   The power of the EFT is then manifest as one can, for the first time, easily include finite size effects in radiation reaction, thus generalizing this famous equation~\cite{toappearGLR}.

\begin{figure}[!t]
\centerline{\scalebox{0.3}{\includegraphics{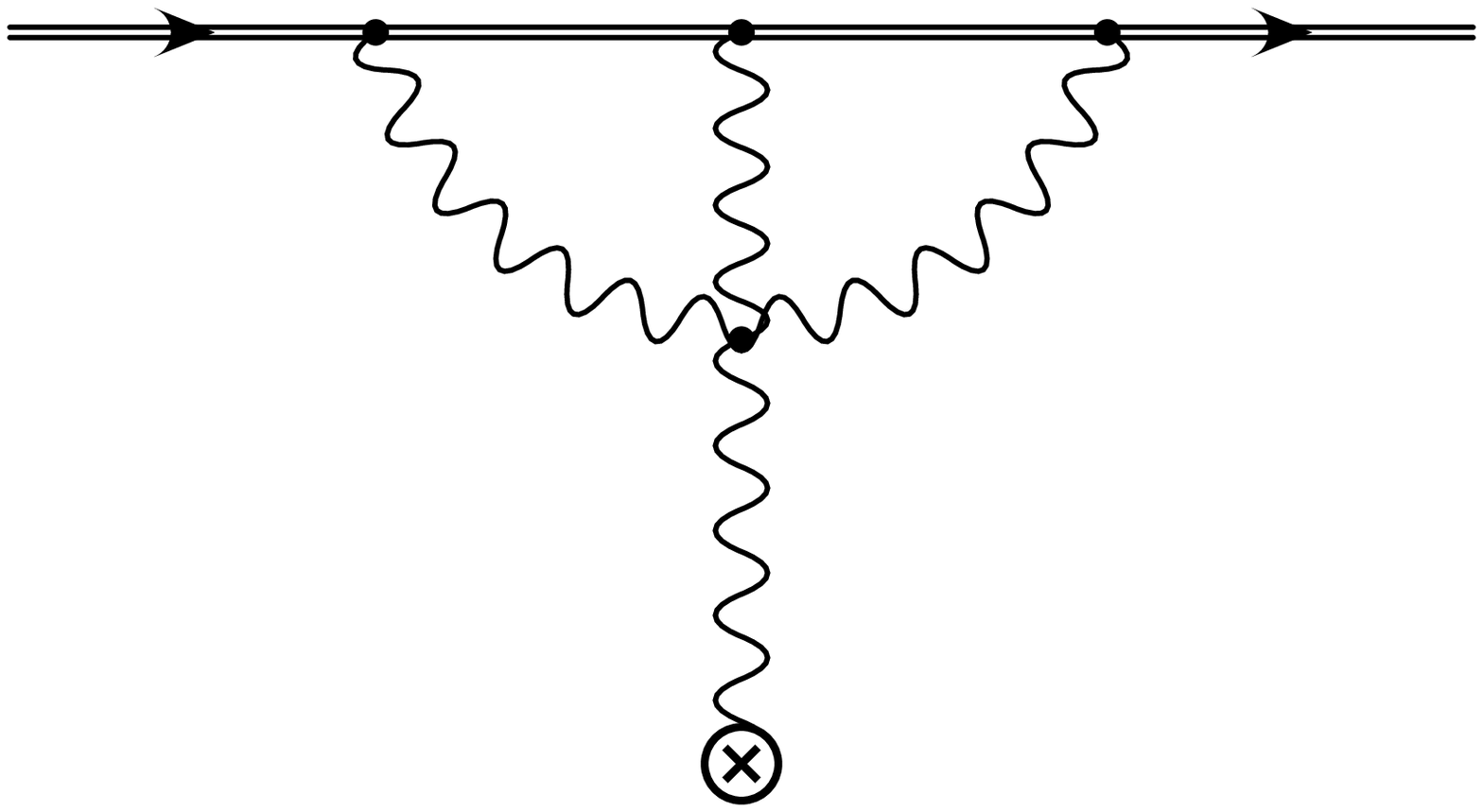}}}
\vskip-0.3cm
\caption[1]{A singular diagram that leads to running of the one body theory between the scales $r$ anad $r/v$. The double line serves as a reminder that it is an effective one body theory.}
\label{subD}
\end{figure}

Using the EFT procedure, it is possible to carry out calculations of post-Newtonian effects to arbitrary order in the expansion parameter $v$.   Short distance divergences due to the point particle limit arise at every order in perturbation.   These are handled by the usual regularization/renormalization program of quantum field theory\cite{TASI}.   For low orders in $v$, such divergences scale like powers of the cutoff, and are therefore not physical (they can be absorbed into mass renormalization).    At order $v^6$, however, one begins to encounter logarithmic divergences, which lead to non-trivial renormalization group flows in the Wilson coefficients of the EFT~\cite{GnR1}.     For objects that are perfectly spherical, such divergences are renormalized by worldline operators of the form $\int d\tau R$, $\int d\tau R_{\mu\nu} v^\mu v^\nu$, which can be removed by field redefinitions of the metric, provided the black hole binary is in a background vacuum spacetime.   They are therefore not physical.    However, if the object has a non-vanishing quadrupole moment $Q_{ij}$ then there is a logarithmic divergence that requires a counterterm operator of the form $\int d\tau \ddot Q_{ij} R^{\mu i \nu j} v_\mu v_\nu$, leading to classical renormalization group logarithms in physical observables\cite{GW}.   For instance, the composite object obtained by integrating out the orbital scale $r$ in the black hole binary has a $Q_{ij}\neq 0$, and so one expects, by the power counting of the EFT, terms of the form $v^6 \log v$ relative to the leading order quadrupole power loss which are calculable by renormalization group methods~\cite{toappearGR}.   A typical Feynman diagram that contributes to the renormalization group equations at order $v^6$ is given in Fig.~\ref{subD}.   

The tower of theories is easily generalized to include more degrees of freedom. The two examples that have been explored so far are the inclusion of spin~\cite{porto} and of the modes responsible for absorption by the black hole horizon~\cite{GnR2}.  To include worldline degrees of freedom  one must first introduce a local frame basis $(v^\mu,e_\alpha^\mu)$.   A generalized angular velocity is
then defined via
\begin{equation}
\frac{D}{D\lambda} e^\mu_I(\lambda)=\Omega^{\mu \nu}e_{\nu I}.
\end{equation}
and the spin $S_{\mu\nu}$ is the conjugate variable to $\Omega^{\mu\nu}$.  To study spin dynamics, one can write down an action, which ignoring finite size effects (which can be treated systematically by the addition of non-minimal operators~\cite{porto}) is fixed by reparameterization invariance to be 
\begin{equation}
\label{eq:spinS}
S=-\int d\lambda \frac{1}{2} S^{\mu \nu}(x(\lambda),\Omega(\lambda)) \Omega_{\mu \nu}.
\end{equation}
Starting from this action, it is straightforward to integrate out the modes at distances shorter than the two-body scale $r$,  and determine the spin-spin generalization of the EIH Lagrangian, thus closing a chapter that begin nearly a century ago.
This rather lengthy result \cite{PR} is an important correction to the dynamics of binary inspirals seen by LIGO and LISA  follows from the calculation of a handful of Feynman diagrams given the action Eq.~(\ref{eq:spinS}).   Other new results for spinning finite size objects such as finite size radiation reaction effects and the $v^6$ radiation power spectrum also follow simply from this formalism

The inclusion of horizon absorption leads to some beautiful new results that relate to recent ideas
regarding the nature of the the black hole horizon.   It has been proposed that in order to account for black hole entropy, it is necessary to include new horizon localized degrees of freedom.  From an effective field theory point of view, there is no choice but include new degrees of freedom on the horizon once the effects of dissipation are included, since dissipation implies the existence of multiple gapless modes.   Although we do not know the precise nature of these degrees of freedom, it is possible to use the $SO(3)$ symmetry of the black hole solution to classify the spectrum of possible composite operator in the worldline theory.   Using $SO(3)$, the couplings of the horizon modes to gravity is of the form
\beq
\label{wlops}
S_{int}= - \int d\tau Q^E_{ab}(\tau) E^{ab} - \int d\tau Q^B_{ab}(\tau) B^{ab}, 
\eeq
where $Q^E_{ab}$ and $Q^B_{ab}$ are composite operators built from the horizon degrees of freedom.   

\begin{figure}[!t]
\centerline{\scalebox{0.50}{\includegraphics{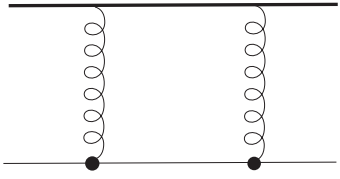}}}
\vskip-0.3cm
\caption[1]{The leading contribution to the dissipative potential . The dots correspond to an insertion of the worldline operators in Eq.~(\ref{wlops}).}
\label{BO0} 
\end{figure}

It is possible to express physical observables in terms of correlators of the horizon operators.   For example the effects of black hole absorption on the dynamics of a binary can be computed from the imaginary part of the box diagram in Fig.~\ref{BO0}, whose value is given in terms of the two-point function
\beq
\int d\tau e^{-i\omega\tau} \langle \Omega| T Q^{E,B}_{ab}(\tau)Q^{E,B}_{cd}(0)|\Omega \rangle=-{i\over 2}\left[\delta_{ac}\delta_{bd}+\delta_{ad}\delta_{bc}-{2\over 3}\delta_{ab} \delta{cd}\right] F(\omega),
\eeq
where $\Omega$ is the vacuum of field theory that describes the horizon modes.   Although we do not know the precise form of the theory that would allow us to calculate this correlation function, the same quantity arises in the expression for the graviton absorption cross section, $\sigma_{abs}(\omega)=\omega^3\mbox{Im} F(\omega)/2 m^2_{Pl}$, that controls the greybody corrections to the Hawking spectrum~\cite{GnR2}.     We therefore find the relation
\beq
\frac{dP_{abs}}{d\omega}=-\frac{1}{T}\frac{G_N}{64\pi^2} \sum_{a\neq b} \frac{\sigma_{abs}^b(\omega)}{\omega^2} m_a^2 |q^a_{ij}(\omega)|^2
\eeq
where $a,b=1,2$ and 
\beq
q^a_{ij}(\omega)=\int dt e^{-i\omega t} \partial_j \partial_j |{\bf x}_{12}|.
\eeq

This is a new relation between the observable $dP/d\omega$ and the cross section $\sigma(\omega)$ which is valied for objects of arbitrary internal composition.     This relation is particularly interesting for a black hole, in which case the above formula can be interpreted as an experimental probe of the unitary theory that is dual to the black hole, a purely gravitational object.   There are several proposals as to the nature of this theory\cite{horth}.    Although knowing the underlying theory would enable one to calculate the dissipative power to all orders, this is not needed to obtain results at leading order in the non-relativistic limit, given that the graviton absorption cross section has been calculated~\cite{staro} by purely gravitational methods.    Plugging the result of~\cite{staro} into our master formula we find the time averaged power loss due to absorption in a black hole binary is
\beq
P_{dis}= \frac{32}{5} G_N^7 m^6 \mu^2 \left \langle \frac{{\bf v}^2}{|{\bf x}|^8}+2 \frac{({\bf x}\cdot {\bf v})^2}{|{\bf x}|^{10}} \right \rangle,
\eeq
where we work in CM coordinates, and $m=m_1+m_2$, $\mu = m_1 m_2/m$.

Finally, it is interesting to note the similarities between our methods and those of the AdS/CFT correspondence.   Recall that AdS/CFT relates the correlators of the field theory on a brane worldlvolume to quantites in semi-classical gravity (in the large $N$ limit).   Likewise, in our case the correlators of a worldline field theory can be expressed in terms of a classical gravity observable, namely the cross section for gravitational wave absorption.  In our formalism this follows directly, in a model-independent way, from the basic tenets of the effective field theory philosophy.   This is in contrast to the string theoretic realizations of the AdS/CFT correspondence, where both sides of the duality are understood and a precise map between the gravitational and non-gravitational theories can be established.  Nevertheless, it is rather surprising to think that this set of ideas, not usually associated to empirically verifiable physics could in fact be a powerful tool in future gravitational wave experiments.

\end{document}